\documentclass{elsart1p}

\usepackage{graphicx}
\usepackage{amssymb}

\def\be{\begin{equation}}
\def\ee{\end{equation}}
\def\bea{\begin{eqnarray}}
\def\eea{\end{eqnarray}}
\def\bear{\begin{array}}
\def\ear{\end{array}}
\def\bfig{\begin{figure}}
\def\efig{\end{figure}}
\def\bcen{\begin{center}}
\def\ecen{\end{center}}
\def\bi{\begin{itemize}}
\def\ei{\end{itemize}}

\def\raw{\rightarrow}

\begin{document}

\begin{frontmatter}


\title{Coherent pion production in neutrino-nucleus collisions}
\author[UV]{L. Alvarez-Ruso},
\ead{Luis.Alvarez@ific.uv.es}
\author[UV]{L. S. Geng},
\author[Nara]{S. Hirenzaki},
\author[UV]{M. J. Vicente Vacas}
\address[UV]{Departamento de F\'{\i}sica Te\'orica and IFIC, Universidad de Valencia - CSIC, Valencia, Spain}
\address[Nara]{Department of Physics, Nara Women's University, Nara, Japan} 

\begin{abstract}
We have investigated the neutrino induced coherent pion production reaction 
 at low and intermediate energies including pion, nucleon and $\Delta(1232)$ resonance. Medium effects in the production mechanism and the distortion of the pion wave function are included in the calculation. We find a strong reduction of the cross section due to these effects and also substantial modifications in the 
energy distributions of the final pion. The dependence  on the atomic number is 
 discussed.
\end{abstract}

\end{frontmatter}

The study of weak coherent pion production in charged current (CC) 
$\nu + A \raw l^- + \pi^+ + A$ and neutral current (NC)  $\nu + A \raw \nu  + \pi^0 + A$ processes  is in the agenda of several current and 
future experiments. The NC reaction is particularly important as a source of background in $\nu_e$ 
appearance oscillation  experiments. Earlier measurements on different nuclei are available at 
$E_\nu > 2$~GeV (see Ref.~\cite{Zeller:2003ey} for a compilation). These data are well described with models 
based on PCAC extrapolated to $q^2 \neq 0$~\cite{Rein:1982pf,Paschos:2005km}. 
The only nuclear medium effect considered, pion distortion,  is  taken into account by factorizing 
the pion-nucleus elastic cross section (c.s.). However, at lower energies, K2K ($\langle E_\nu \rangle=1.3$~GeV), finds a  significant deficit of muons in the forward scattering events with respect to the 
simulations, which has been interpreted as due to an overestimation of CC coherent $\pi^+$ production~\cite{Hasegawa:2005td}. Meanwhile, the MiniBooNE NC $\pi^0$ data set is under analysis. 

We have performed a theoretical study of neutrino induced coherent pion~\cite{AlvarezRuso:2007tt,AlvarezRuso:2007it} production extending and improving 
the calculations of Kelkar {\it et al.}~\cite{Kelkar:1996iv} and Singh {\it et al.}~\cite{Singh:2006bm}. The relativistic amplitude is built in terms of the relevant hadronic degrees of freedom.  
Besides the dominant direct $\Delta$ excitation, this model includes the 
crossed $\Delta$ and nucleon-pole terms, all of them represented    
in Fig.~\ref{Fig1}. There are other contributions allowed by chiral symmetry~\cite{Hernandez:2007qq} but they 
cancel for isospin symmetric nuclei, so we neglect them. 
\bfig[ht]
\begin{minipage}{.23\linewidth}
\includegraphics[width=0.9\textwidth]{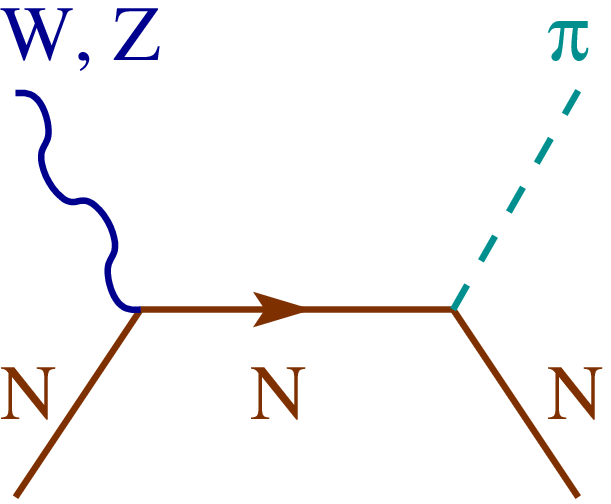}
\end{minipage}
\hfill
\begin{minipage}{.23\linewidth}
\includegraphics[width=0.9\textwidth]{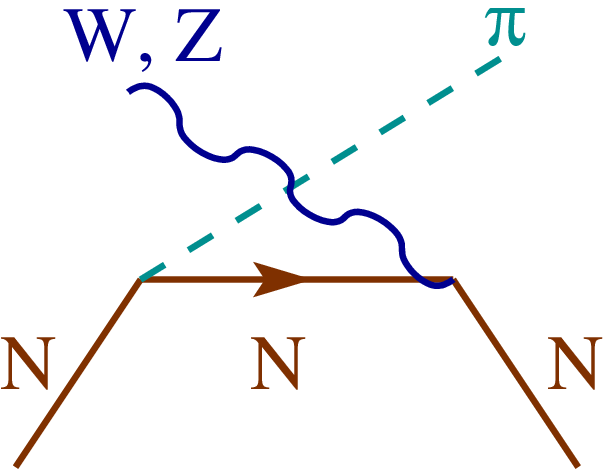}
\end{minipage}
\hfill
\begin{minipage}{.23\linewidth}
\includegraphics[width=0.9\textwidth]{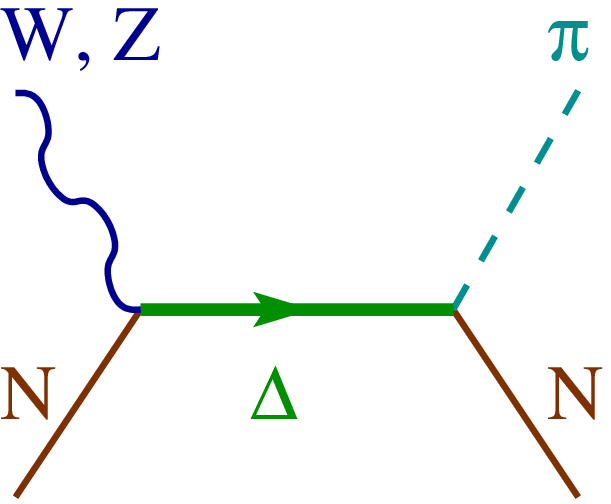}
\end{minipage}
\hfill
\begin{minipage}{.23\linewidth}
\includegraphics[width=0.9\textwidth]{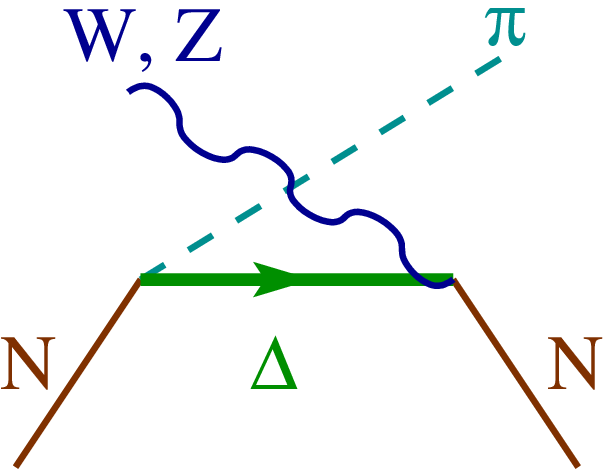}
\end{minipage}
\caption{Elementary mechanisms for coherent pion production on isospin-symmetric nuclei} 
\label{Fig1}
\efig

The cross section is given by 
\begin{equation}
\frac{d \sigma}{d \Omega' d E' d \Omega_\pi} = \frac{1}{8}
\frac{|\vec{k}'| |\vec{p}_\pi|} {|\vec{k}|} \frac{1}{(2 \pi)^5} \,
{{G^2}\over{2}}| l_{\mu} J^{\mu}_{A} |^2\,,
\end{equation}
where $(E_\nu,\vec{k})$, $(E',\vec{k}')$ and $(\omega_\pi,\vec{p}_\pi)$ are the 4-momenta of the incoming neutrino, outgoing lepton and 
outgoing pion respectively; $l_{\mu}= \bar{u}_{l'} (k') \gamma_{\mu} (1 -\gamma_{5}) u_{\nu}(k)\,$ 
is the standard leptonic current. The nuclear current $J^{\mu}_{A}$ is obtained as the coherent sum over all nucleons, performed in the framework of a local Fermi gas. Detailed expressions of $J^{\mu}_{A}$ for all four mechanisms can be found in Ref.~\cite{AlvarezRuso:2007it}. The single-nucleon 
contributions to the current are parametrized in terms of vector and axial form factors (FF). The vector FF are related to the electromagnetic ones and can be extracted from electron scattering data. The axial FF are usually constrained by means of PCAC. 

As can be seen from the comparison of the two dashed lines in Fig.~\ref{Fig2}, the direct $\Delta$ excitation mechanism accounts for most of the c.s. at $E_\nu \sim 1$~GeV. Actually, the crossed $\Delta$ amplitude is very small, and there is a cancellation between the direct and crossed pole-nucleon ones. This implies that, with high accuracy, $\sigma \propto \left[ C_5^A(0) \right]^2$ because this is the only FF that contributes at $q^2=0$, where most of the strength of the coherent reaction is concentrated. Here we use $C_5^A(0)=1.2$~\cite{AlvarezRuso:1998hi}, in agreement with the off-diagonal Goldberger-Treiman relation, but a recent phenomenological analysis~\cite{Hernandez:2007qq} of ANL data, as well as lattice~\cite{Alexandrou:2006mc} and quark model~\cite{BarquillaCano:2007yk} calculations suggest a smaller value.
\bfig[ht!]
\begin{minipage}{.49\linewidth}
\bcen
\includegraphics[width=0.99\textwidth]{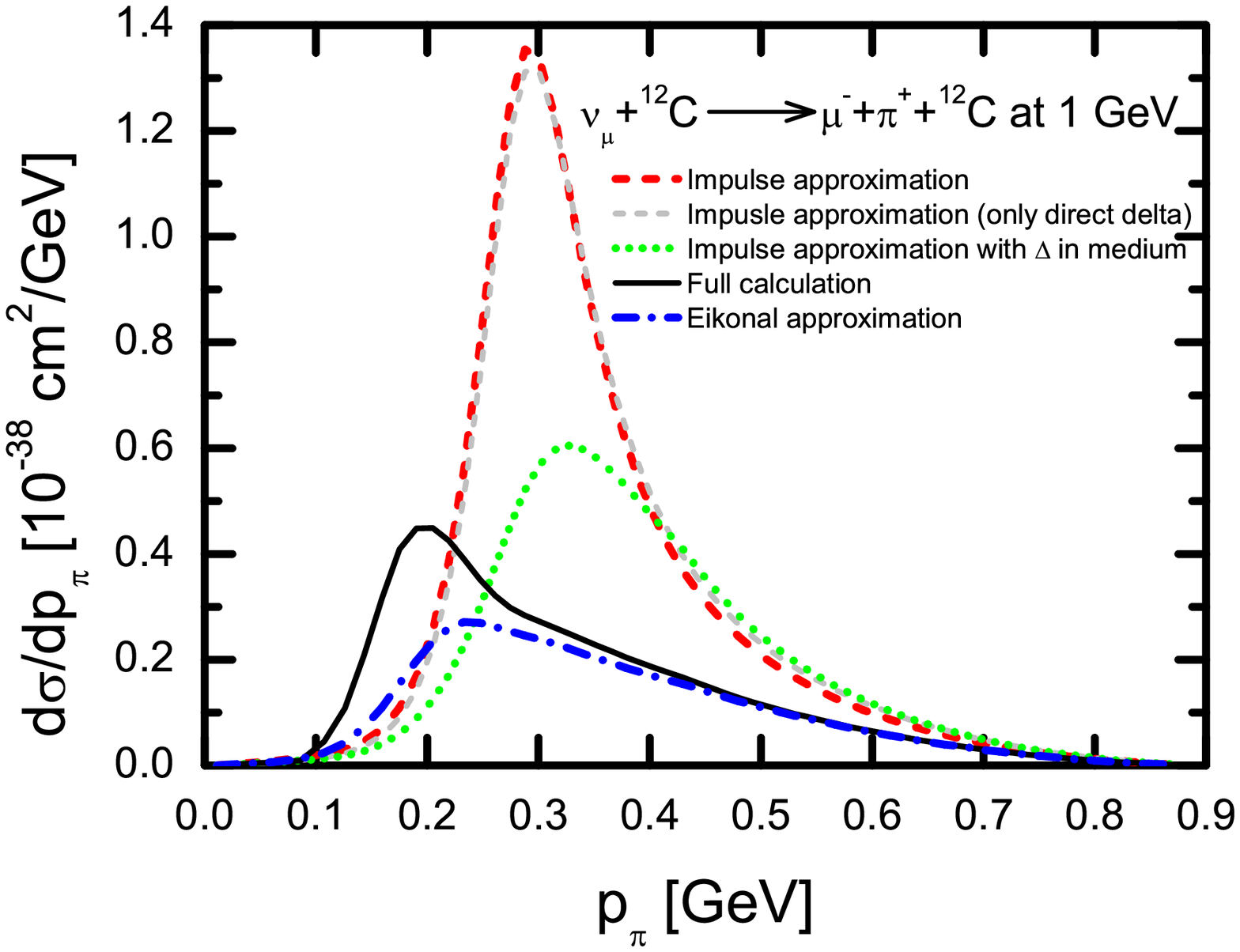}
\caption{Pion momentum distribution.} 
\label{Fig2}
\ecen
\end{minipage}
\hfill
\begin{minipage}{.49\linewidth}
\bcen
\includegraphics[width=\textwidth]{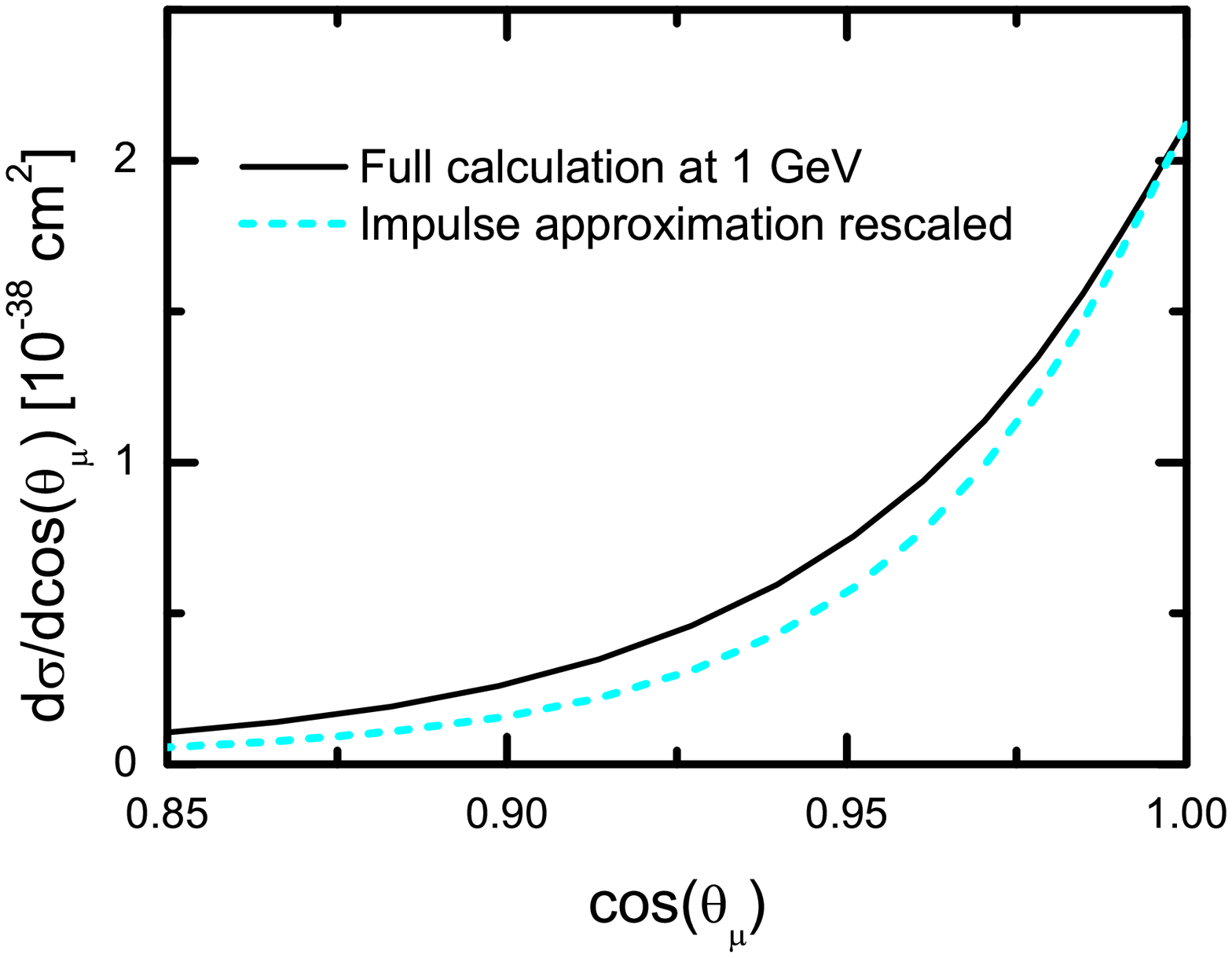}
\caption{Muon angular distribution.}
\label{Fig3}
\ecen
\end{minipage}
\efig

The $\Delta$ properties are strongly modified inside the nuclear medium. We take this into account by adding a density dependent selfenergy and modifying the free width in the $\Delta$ propagator. 
This modification reduces  the c.s. by around 35~\%. 
Unlike the incoherent case, where a semiclassical treatment of pion final state 
interactions is required~\cite{Leitner:2006ww}, now the nucleus remains in its ground state, so a  realistic quantum 
treatment of pion distortion can be achieved by solving the Klein-Gordon equation 
with a microscopic optical potential $\hat{V}_{opt}$~\cite{GarciaRecio:1989xa,Nieves:1991ye} 
based on the $\Delta$-hole model.  Pion distortion further  decreases 
the c.s. and moves the peak to lower energies. This reflects the presence of a strongly absorptive part in $\hat{V}_{opt}$ around the $\Delta$ peak. The eikonal approximation fails at $p_\pi < 400$~MeV$/c$, where a better treatment of the pion wave function is required. The angular distribution of the emitted muons remains relatively unaffected by these nuclear corrections, as can be seen in Fig.~\ref{Fig3} where we compare our model with the impulse approximation, rescaled to match the full calculation at zero degrees.

We have also investigated the dependence of the c.s. on the atomic number. In the impulse approximation, taking into account the isospin factors, one finds that the amplitude is proportional to an effective number of participants defined as P=Z+N/3; here Z and N are the number of protons and neutrons respectively. This could suggest a quadratic dependence of the cross section on P. However, our full model exhibits a different trend. The P dependence is quenched for two reasons. First, the strong pion absorption forces the reaction to be peripheral. Second, the nuclear form factor is narrower for heavy nuclei, and reduces more the contribution from high momentum transfers.
\bfig[h!]
\bcen
\includegraphics[width=0.44\textwidth]{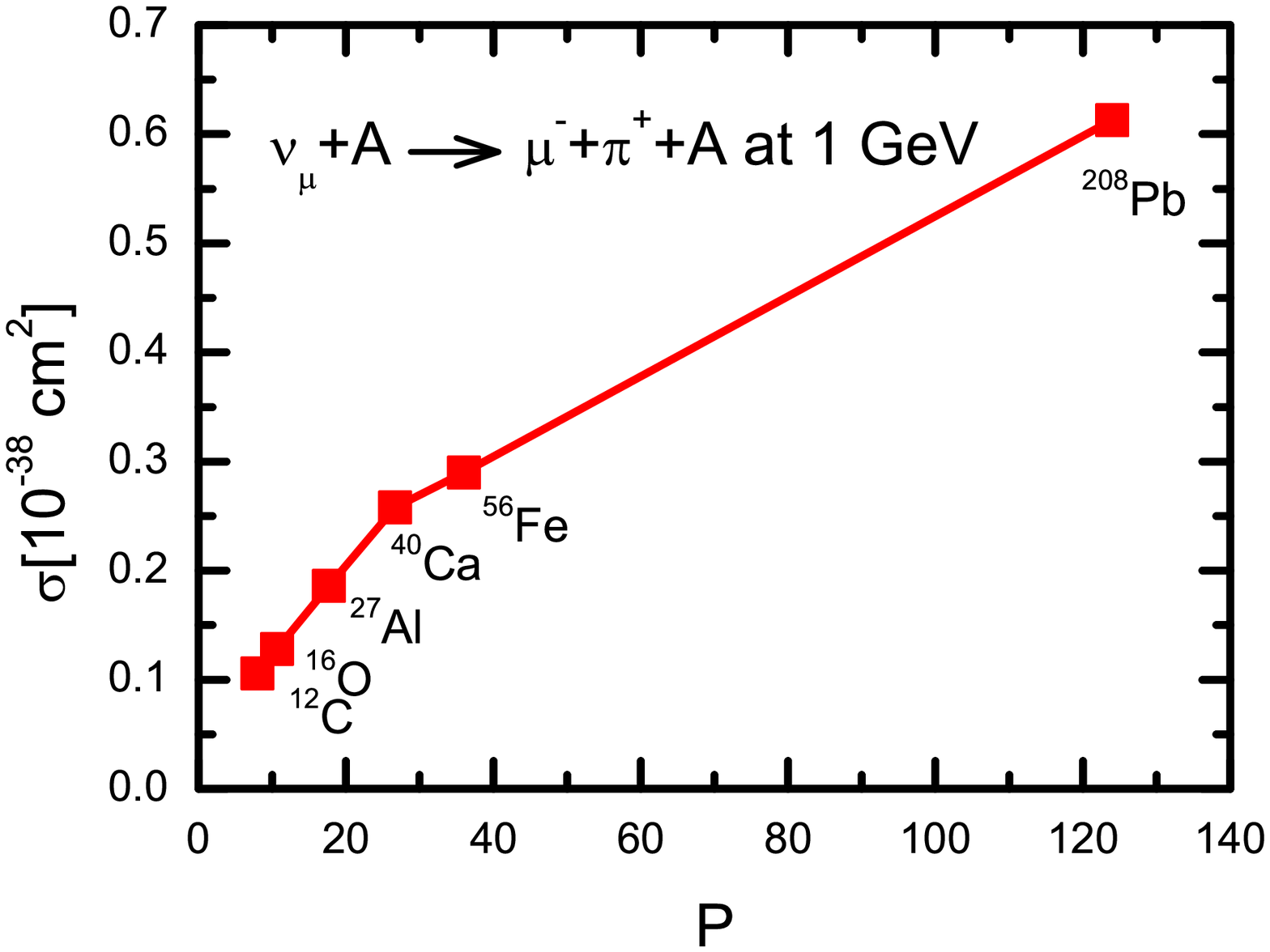}
\ecen
\caption{Total cross section as a function of the effective number of participants P=Z+N/3}
\efig




\end{document}